# Suppression of FM-to-AM conversion in third-harmonic generation at the retracing point of a crystal

Yisheng Yang,[1,2,*] Bin Feng,[2] Wei Han,[2] Wanguo Zheng,[2] Fuquan Li,[2] and Jichun Tan[1]

[1]*College of Science, National University of Defense Technology,*

*Changsha 410073, China*

[2]*Research Center of Laser Fusion, China Academy of Engineering Physics,*

*Mianyang 621900, China*

[*]*Corresponding author: ysyang@nudt.edu.cn*

FM-to-AM conversion can cause many negative effects (e.g., reducing of margin against damage to the optics, and etc.) on performances of third-harmonic conversion system. In this letter, the FM-to-AM conversion effect in third-harmonic generation is investigated both at and away from the retracing point of type-II KDP crystal. Obtained results indicate that the FM-to-AM conversion can be suppressed effectively when the crystal works at the retracing point.

In broadband third-harmonic generation (THG), phase modulation (i.e., frequency modulation, FM) of input field can be converted to intensity modulation (i.e., amplitude modulation, AM) of output harmonic field, which is called as the FM-to-AM conversion [1]. Generally, this conversion adversely affects the laser performance, and needs to be prevented. The resulting AM can cause damage to optical elements slightly outside the near-field (e.g., image relay optics), or may lead to some higher-order nonlinear effects due to instantaneous ultra-high intensity in process [2, 3].

With the purpose of weakening or averaging the resulting AM, some specific approaches were proposed. For example, plane gratings (NIF configuration) or focusing gratings (LMJ configuration) are introduced in the SSD (smoothing by spectral dispersion) systems, and have been demonstrated to be able to reduce the AM strongly [4, 5]. As well, adding a second mixer to the baseline configuration (i.e., one doubler followed by one mixer) can lower the resulting AM to some certain extent [6]. In this letter, the issue of FM-to-AM conversion in the THG process is investigated from the viewpoint of retracing properties of nonlinear crystal, and a criterion to suppress the FM-to-AM conversion purposely is also presented.

Retracing point refers to the point or wavelength of a phase-matching curve (phase-matching angle $\theta_{PM}$ versus wavelength λ) at which slope vanishes, as the point R shown in Fig. 1. It has the characteristic of zero group-velocity-mismatch (i.e., the first-order wavelength-sensitivity of phase-mismatch $\partial(\Delta K)/\partial\lambda$ equals to zero at the retracing point.) [7]. According to Ref. [8], the THG retracing point corresponds to a special group-velocity-matching (GVM) relationship of $3/\upsilon_3 = 1/\upsilon_1 + 2/\upsilon_2$, where $\upsilon_1$, $\upsilon_2$, and $\upsilon_3$ are the group-velocities of 1ω, 2ω, and 3ω pulses propagating in crystal, respectively, as presented in Fig. 1.

For third-harmonic generation of sinusoidally phase-modulated Gaussian pulses, the input amplitude profile $E(t)$ can be written in the following form:

$$E(t) = E_0 \exp\left(-\frac{t^2}{2T_0^2}\right)\exp\left[i\sigma\sin(2\pi\Omega t)\right], \tag{1}$$

where $E_0$ is original amplitude which is equal to $\sqrt{2I_0/nc\varepsilon_0}$, and $T_0$, σ, and Ω are the pulse duration (i.e., half-width at 1/e intensity), modulation amplitude, and modulation frequency, respectively. Optical spectrum of the sinusoidal phase modulation $\exp\left[i\sigma\sin(2\pi\Omega t)\right]$ can be given by a sum over Bessel functions (i.e., $\sum_{n=-\infty}^{\infty} J_n(\sigma)\delta(f-n\Omega)$), which has a frequency bandwidth (FWHM) of about 2σΩ [9].

Based on the split-step Fourier transform and the fourth-order Runge-Kutta algorithm, the THG process in type-II KDP crystal is numerically investigated in two cases: at the retracing point and away from the retracing point (i.e., group-velocity matched and group-velocity mismatched cases). In accordance with Fig. 1, we refer to the above two cases as "line n" and "line m", respectively. For better comparison of conversion performances in above two cases, both input 1ω wavelengths in simulation are set to be 1.053 μm. For the "line m" case, group-velocities of three waves can be easily obtained through the Sellmeier equations of KDP crystal (i.e., $\upsilon_1 = 2.02\times10^8$ m/s, $\upsilon_2 = 1.94\times10^8$ m/s, and $\upsilon_3 = 1.92\times10^8$ m/s) [10]. While for the other case, just substitute the value of $\upsilon_3$ by $1.97\times10^8$ m/s to satisfy the special GVM relationship $3/\upsilon_3 = 1/\upsilon_1 + 2/\upsilon_2$. Values of other basic parameters of the input 1ω and 2ω pulses in simulation are given in Table 1.

Figs. 2 and 3 compare the intensity temporal profiles of the three output harmonics in above two cases. We find that, the intensity profiles in Fig. 2 are severely modulated while curves of Fig. 3 appear to be much smoother, which implies that the FM-to-AM conversion in the THG process can be suppressed effectively provided the nonlinear crystal works at its THG retracing point. Such suppression of the FM-to-AM conversion can be interpreted via the spectrum contrast defined by $\tilde{I}_{in}(f)/\tilde{I}_{out}(f)$, where $\tilde{I}_{in}(f)$ and $\tilde{I}_{out}(f)$ are the spectrum intensity of input and output fields, respectively. Fig. 4 gives the spectrum contrast of 1ω pulses in the two cases. It can be seen that, working at the retracing point, the shape of spectrum contrast is symmetrically distributed and centralized while it is asymmetric and dispersed in the case of working away from the retracing point. According to the definition $\tilde{I}_{in}(f)/\tilde{I}_{out}(f)$, value of the spectrum contrast is inversely proportional to the output (i.e., residual for 1ω pulse) spectrum intensity, and thereby is proportional to the conversion efficiency. Compared with the chaotic spectrum contrast in the "line m" case, the regular one in the case of "line n" reflects that the optical spectrum of pulse are almost not filtered when the crystal operates at its THG retracing point [4].

To quantitatively compare the FM-to-AM conversion in above two cases, an index α is introduced to evaluate the distortion of resulting AM, which is defined by the following expression [5]:

$$\alpha = 2\frac{I_{max} - I_{min}}{I_{max} + I_{min}}. \tag{2}$$

where $I_{max}$ and $I_{min}$ are the maximum and minimum value of pulse intensity at the time scale of [-$T_0$/100, $T_0$/100], respectively. α ranges from 0 to 2, and is ideally equal to 0 (namely, no intensity modulation). Fig. 5 shows the dependence of α on the bandwidth of 1ω pulse in the cases of "line m" and "line n". Numerical result reveals that the distortion of the resulting AM increases much slower (See the curve marked red star) when the crystal works at the retracing point. In addition, for a given distortion index α, the acceptable fundamental bandwidth can be dramatically extended while working at the retracing point.

In conclusion, the FM-to-AM conversion in the process of THG can be suppressed successfully through adopting crystals with appropriate THG retracing point (or in other words, by satisfying the special group-velocity-matching relationship $3/\upsilon_3 = 1/\upsilon_1 + 2/\upsilon_2$ [8]), which gives theoretical guidance for the suppression of AM in practical applications.

This work was partially supported by the National Natural Science Foundation of China (60708007), Science and Technology Foundation of Chinese State Key Laboratory of High Temperature and Density Plasma Physics (9140C6803010802), and Development Foundation of China Academy of Engineering Physics (2008B0401043).

# Figures and Figure Captions

Fig. 1. Phase-matching curve (blue) and group-velocity curves (red) of type-II KDP mixer. The retracing point R just corresponds with the crossing point O of group-velocity curves at $\lambda_{1\omega} = 1.577\mu m$.

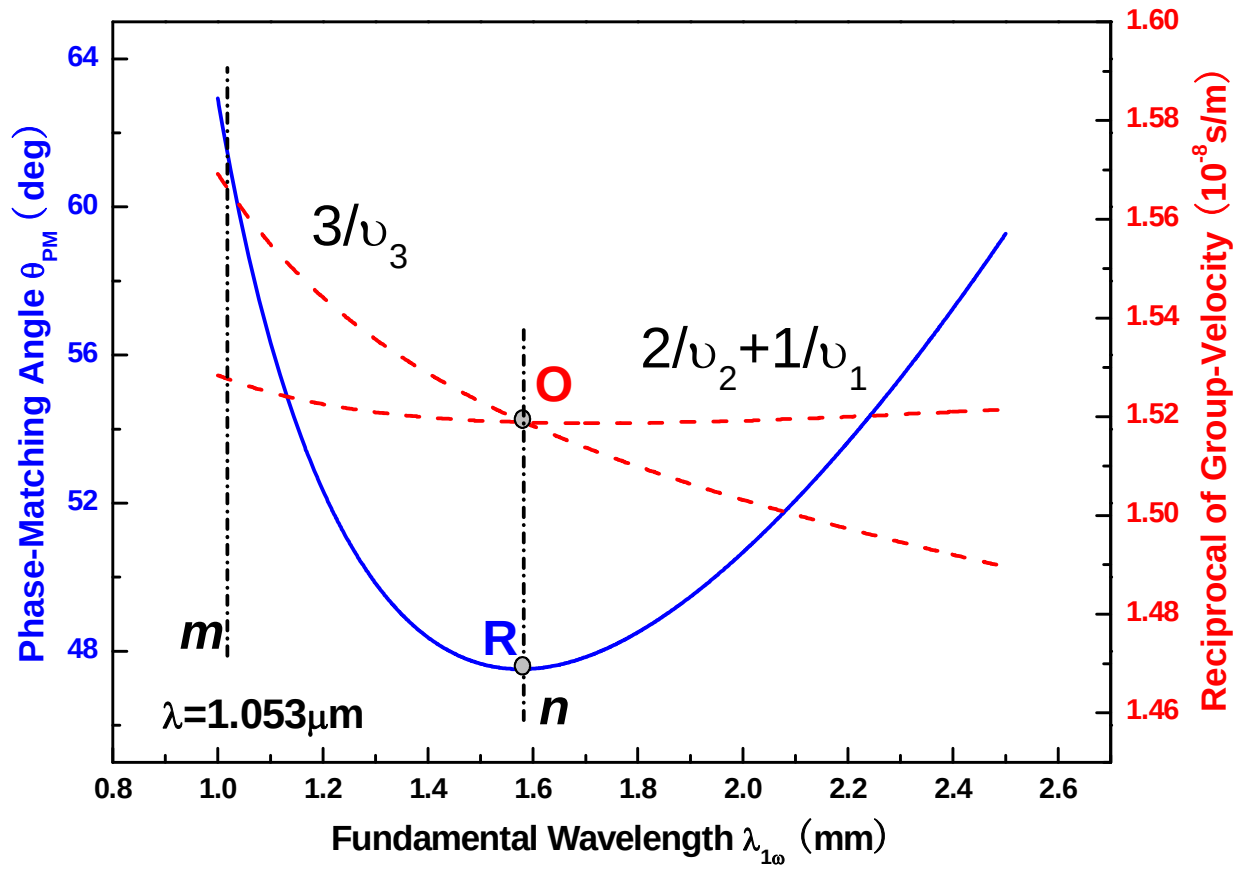

Fig. 2. Temporal profiles of output harmonic pulses in “line m” case. Propagation length in KDP crystal is 6 mm.

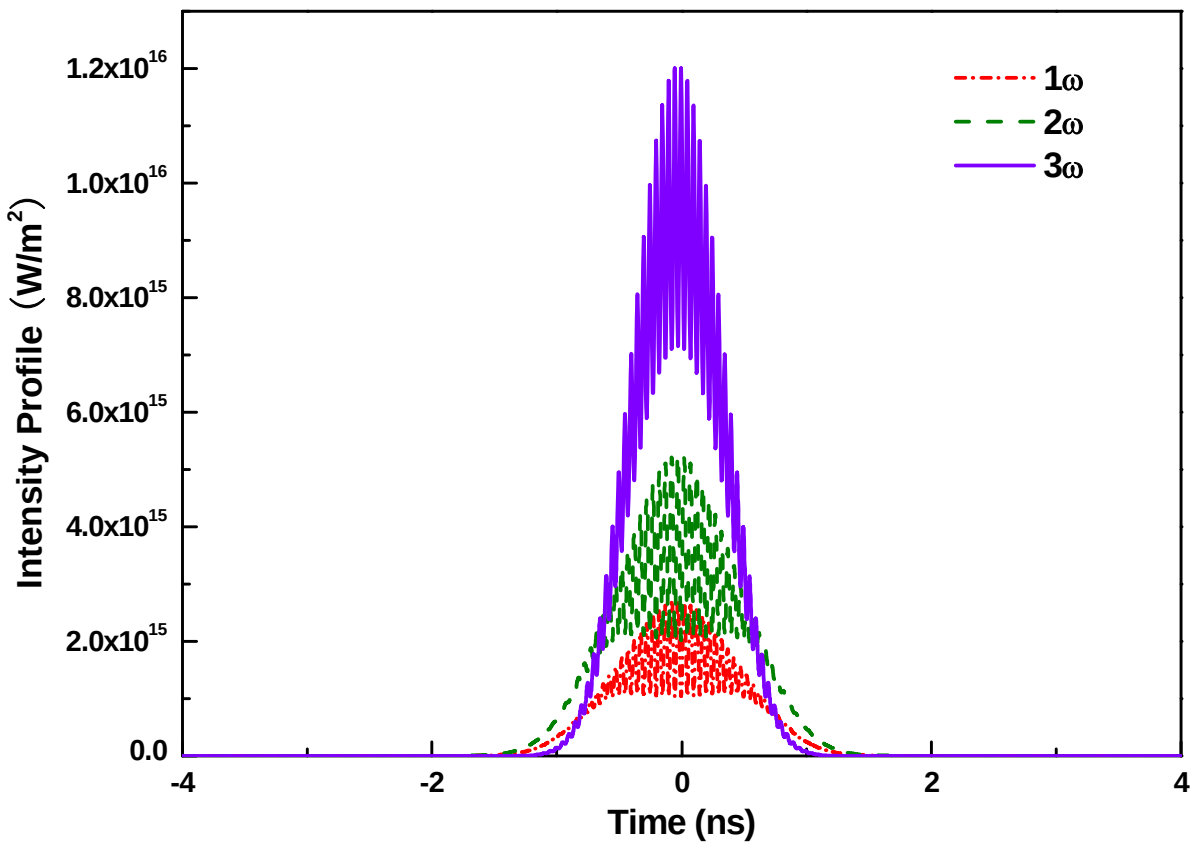

Fig. 3. Temporal profiles of output harmonic pulses in “line n” case. Propagation length in KDP crystal is 16 mm.

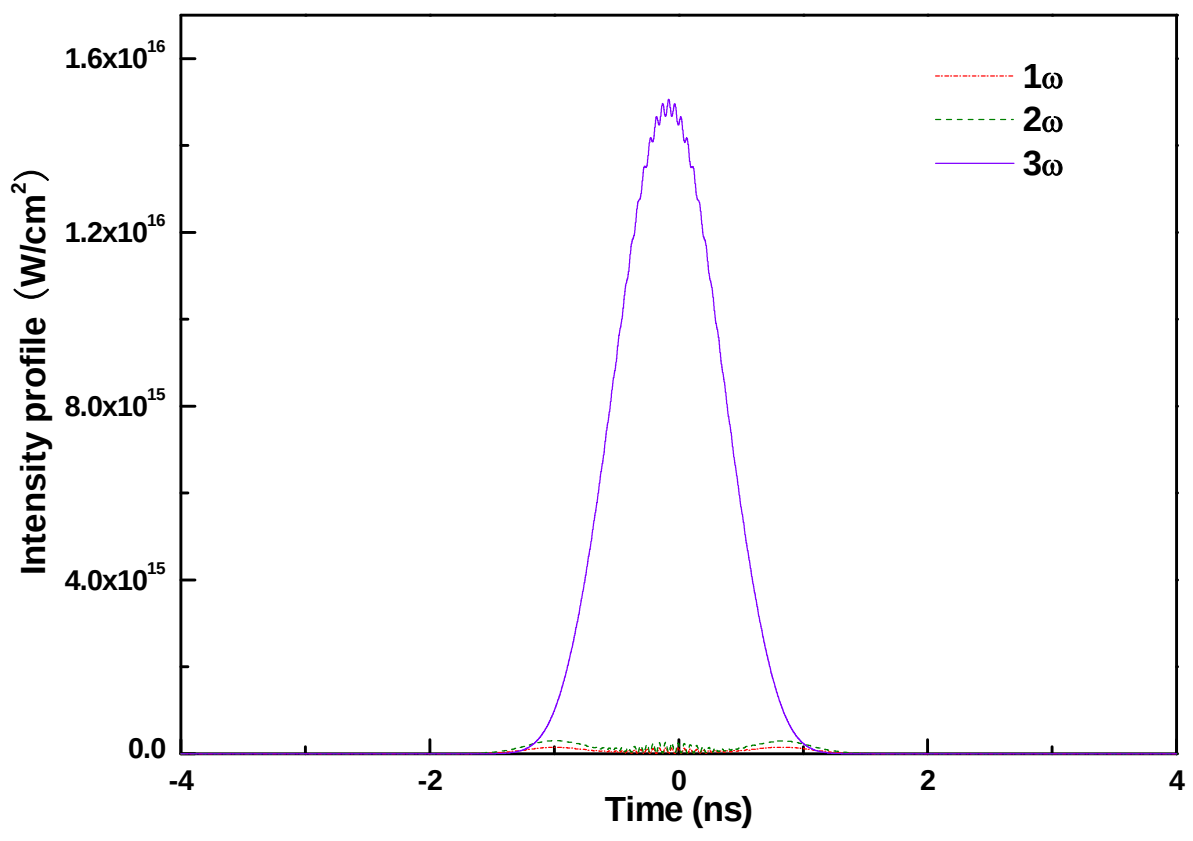

Fig. 4. Spectrum contrast of 1ω pulse in the cases of “line m” and “line n”.

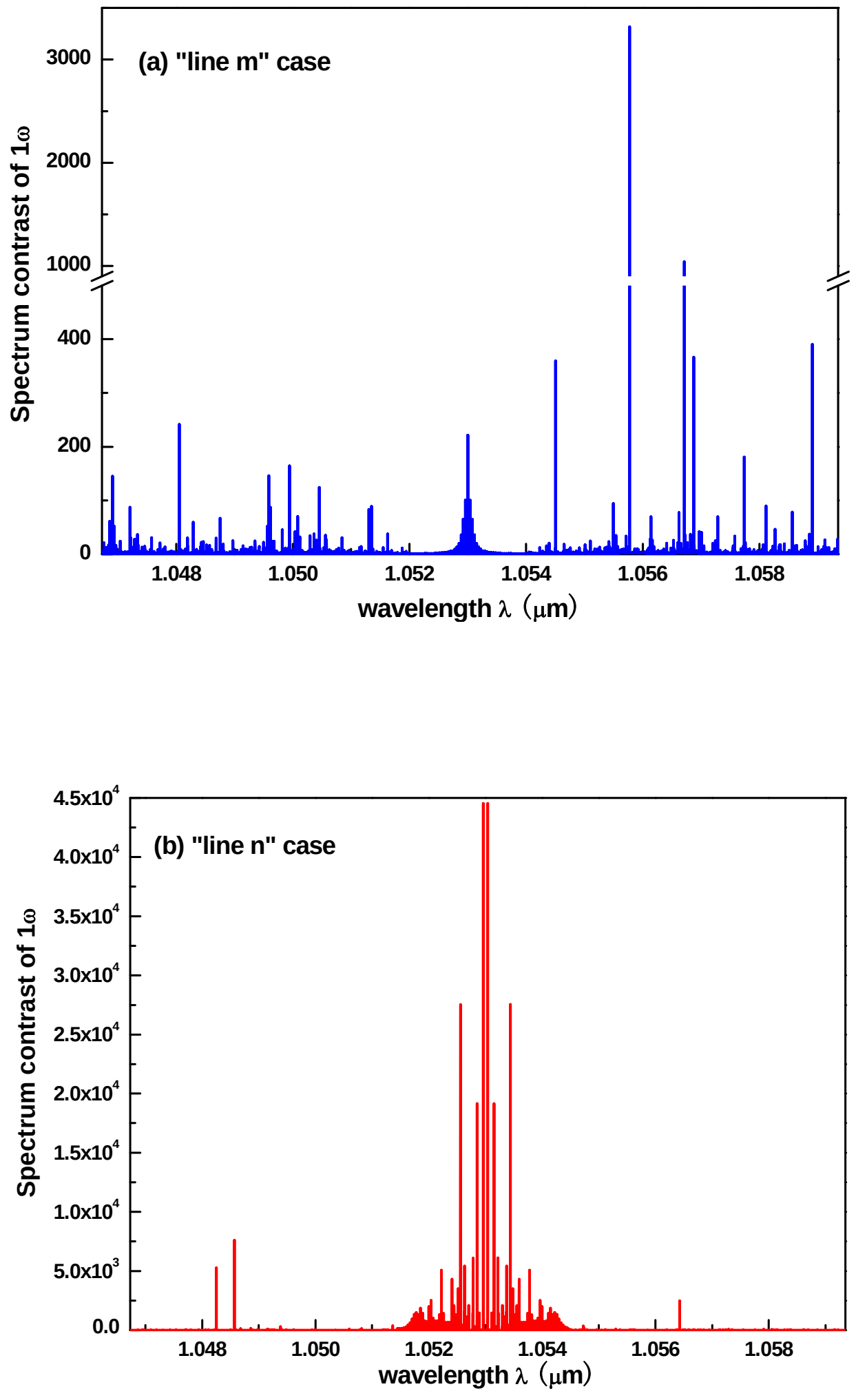

Fig. 5. Dependence of distortion index α on the phase-modulated bandwidth of 1ω pulse in the cases of “line m” and “line n”.

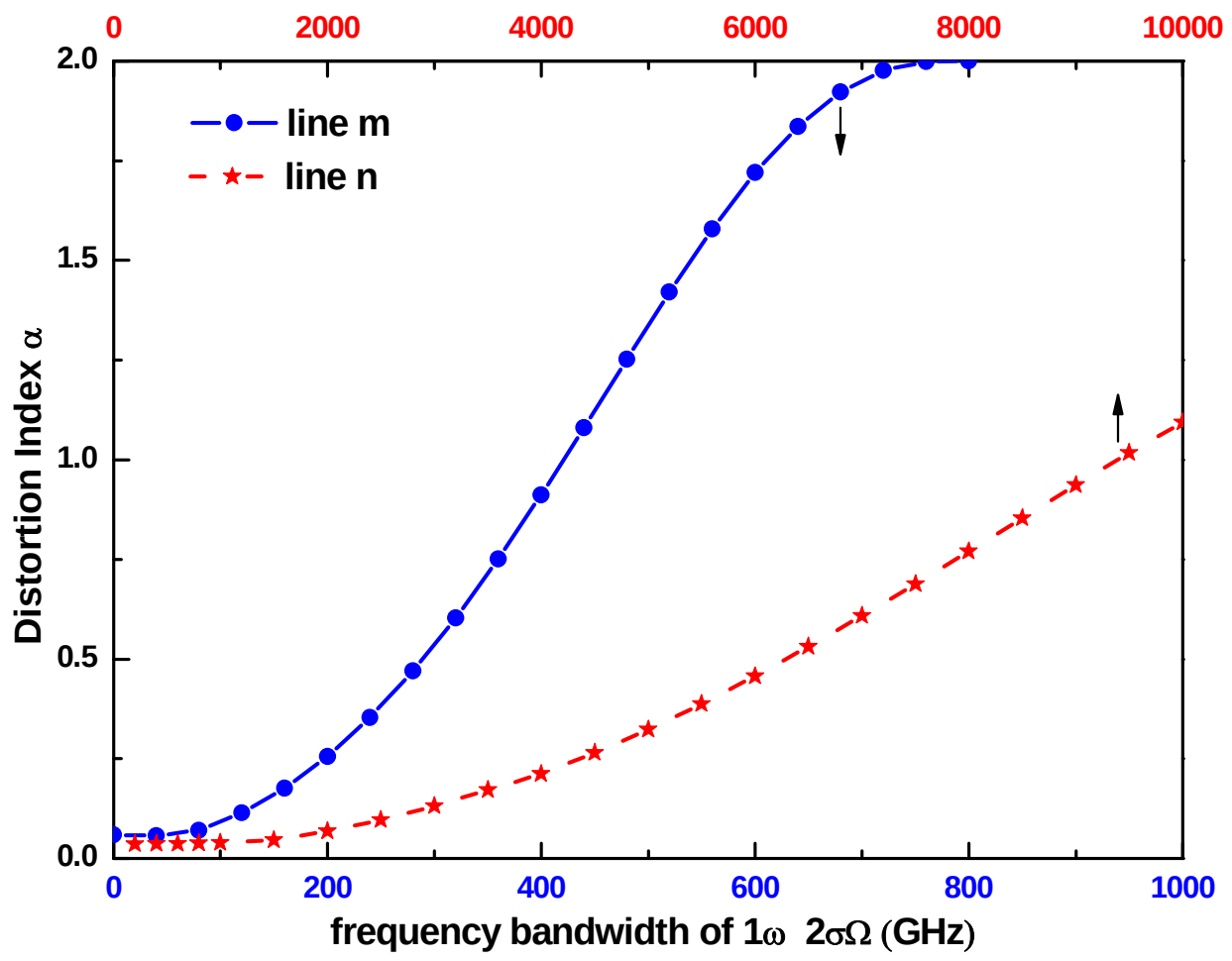

# Tables

Table 1. Values of basic parameters of input 1ω and 2ω pulses

| | Peak Intensity $I_0$ ($GW/cm^2$) | Pulse Duration $T_0$ (ns) | Modulation Amplitude σ | Modulation Frequency Ω (GHz) | Frequency Bandwidth 2σΩ (GHz) |
|---|---|---|---|---|---|
| 1ω/1.053 μm | 1 | 1 | 15 | 10 | 300 |
| 2ω/0.527 μm | 2 | 1 | 30 | 10 | 600 |